\documentclass[10pt,english,aps,showkeys,showpacs,nofootinbib]{revtex4-1}
\usepackage[T1]{fontenc}
\usepackage{textcomp}
\usepackage[utf8]{inputenc}
\usepackage{geometry}
\usepackage{color}
\usepackage{babel}
\usepackage{amsmath}
\usepackage{amssymb}
\usepackage[bookmarks=false,
 breaklinks=false,pdfborder={0 0 1},backref=section,colorlinks=false]
 {hyperref}

\makeatletter
\usepackage{todonotes}
\usepackage[bottom]{footmisc}
\usepackage{graphicx}
\usepackage{dcolumn}
\usepackage{bm}
\usepackage{todonotes}
\usepackage{babel}

\makeatother

\begin{document}
\selectlanguage{english}
\title{Analytical Solution of the PDM-Coulombic Klein-Gordon Oscillator in
Kaluza-Klein Theory}
\author{Abdelmalek Bouzenada}
\email{abdelmalek.bouzenada@univ-tebessa.dz ; abdelmalekbouzenada@gmail.com}

\affiliation{Laboratoire de Physique Appliquée et Théorique~~\\
 Université Larbi-Tébessi-, Tébessa, Algeria}
\author{Abdelmalek Boumali}
\email{boumali.abdelmalek@gmail.com}

\affiliation{Laboratoire de Physique Appliquée et Théorique~~\\
 Université Larbi-Tébessi-, Tébessa, Algeria}
\author{Omar Mustafa }
\email{omar.mustafa@emu.edu.tr }

\affiliation{Department of Physics, Eastern Mediterranean University, ~~\\
 G. Magusa, north Cyprus, Mersin 10 - Turkey}
\author{R. L. L .Vitória}
\email{ricardo.vitoria@pq.cnpq.br ; ricardo-luis91@hotmail.com}

\affiliation{Faculdade de Física, Universidade Federal do Pará, Av. Augusto Corrêa,~~\\
 Guamá, Belém, PA 66075-110, Brazil.}
\author{Marwan Al-Raeei}
\email{mhdm-ra@scs-net.org ; mn41@liv.com}

\affiliation{Faculty of Science,Damascus~~\\
 University,Damascus,Syria}
\date{\today}
\begin{abstract}
In this contribution, the relativistic quantum motions of the position-dependent
mass (PDM) oscillator field with a scalar potential in the context
of the Kaluza-Klein theory is investigated. Through a purely analytical
analysis, the eigensolutions of this system have been obtained. The
results showed that the KGO is influenced not only by curvature, torsion
and the quantum flux associated with the extra dimension, but also
by the possibility of modifying the mass term of the Klein-Gordon
equation.
\end{abstract}
\keywords{Klein-Gordon oscillator, topological defects, Kalluza-Klein Theory,
PDM , Coulomb-Type Potentiel .}
\pacs{03.65.Ge; 03.50.\textminus z; 05.70.Ce ; 03.65.\textminus w ; 04.62.+v;
04.40.\textminus b; 04.20.Gz; 04.20.Jb; 04.20.\textminus q; 03.65.Pm }
\maketitle

\section{Introduction }

One of the fields of investigation in the context of quantum mechanics
is the gravitational effects on the relativistic quantum dynamics
of a particle characterized by well-defined spins, for example, spin-$1/2$
and spin-$0$ particles, described by the Dirac equation and the Klein-Gordon
equation, respectively, both extended to incorporate the curvature
effects intrinsic to the gravitation of the environment considered
as background \cite{bd}.

In particular, spin-0 particles have been heavily investigated subject
to gravitational effects in various types of space-time, for two reasons:
searches for interpretations or extensions of standard models of particle
physics, mainly due to the advent of the Higgs boson with its experimental
evidence and its powerful effect in the Standard Models \cite{ryder},
as well as in the description of particles or quasi-particles in condensed
matter systems in analogous models \cite{kleinert}, such as the case
of phonons \cite{bas}, two-dimensional systems \cite{dantas} and
quantum dynamics in graphene sheets \cite{graff,graff1} with various
defects generated by the process to cut and paste \cite{kat,kat1}.

Spin-$0$ particles have been investigated, for example, in cosmic
string spacetime, subject to central potentials and external fields.
We therefore have studies of particles with position-dependent mass
(PDM) subject to a scalar and electromagnetic Coulomb-type potential,
linear potential, Landau quantization, Aharanov-Bohm effect (ABE)
for bound states \cite{eug} and interacting with the Klein-Gordon
oscillator \cite{bou}. It is noteworthy that the presence of the
cosmic string in the background is associated with the curvature of
spacetime \cite{vil,sc}.

Another ingredient considered in the background in order to extend
the understanding of bosonic spin-$0$ quantum dynamics is the presence
of torsion in the environment \cite{put}. From a mathematical point
of view, torsion in spacetime arises due to the non-symmetry of the
Christoffel symbols, thus producing a tensor field associated with
torsion \cite{kat,kat1,put}. However, torsion can be typified according
to the conservation of the geometric symmetry produced after the "cut
and paste" process \cite{kat,put}. The best-known examples are time-type,
spiral-type, and screw-type dislocations \cite{kat1,put}. In particular,
from the point of view of condensed matter, the screw-like dislocation
is associated with the Burgers vector \cite{kat,put}. Spin-$0$ particle
has been investigated in spacetime with torsion, for example, subjected
to ABE for bound states \cite{ric}, by interacting to electromagnetic
and scalar Coulomb-type potential \cite{ric1}, linear potential and
Klein-Gordon oscillator \cite{ric2}. In addition, there are studies
of scalar particles immersed in the spacetime with a screw-type dislocation
under effects of rotation \cite{ric3,ric4}.

A quantum effect that can be observed of cosmic string and screw-like
dislocation on bound state solutions in quantum particle dynamics:
effect analogous to ABE for bound states \cite{abe,abe1}. That is,
the quantum numbers associated with the angular momentum are modified
by the background topology, generating a kind of "correction" in
the eigenvalues of the angular momentum, which, in turn, provides
an effective angular momentum \cite{valdir,valdir1,boumali3,boumali4,boumali5}.

In particular, ABE for bound states has been investigated in a space-time
with an extra dimension \cite{erico}, known as the Kaluza-Klein theory
(KKT) \cite{kkt,kkt1}. and quantum mechanics is possible, but with
the insertion of extra dimensions. This proposal is defended in the
famous string theory \cite{kkt2}. Although KKT is much simpler than
more modern theories of extra dimensions, it has been widely used
for research into the relativistic quantum dynamics of fundamental
particles \cite{kkt2,bou}. In the case of spin-$0$ particles, we
have studies in systems with PDM \cite{erico1}, Landau quantization
\cite{erico2}, Klein-Gordon oscillator \cite{erico3} and under effects
of rotation \cite{erico4}.

Therefore, in this analysis, we intend to investigate the effects
of curvature \cite{vil} and torsion \cite{put} of space-time incremented
with an extra dimension on an electrically charged scalar quantum
oscillator subjected to a quantum flux analogous to the ABE. Our analysis
will be purely analytical in obtaining the relativistic energy profiles,
as well as in determining the axial eigenfunctions that describe the
oscillations of the electrically charged scalar particle in this non-trivial
spacetime.

The structure of this paper is as follows: in Sec. 2, we make a brief
review on quantum dynamics of a free particle in KKT spacetime with
curvature and torsion, in which we show that the general solution
of the scalar particle is defined through the Bessel function of the
first type redefined in terms of the parameters associated with the
curvature and torsion of KKT space-time; in Sec. 3, we investigated
the effects of central linear potential in the mass of the quantum
oscillator immersed in the KKT space-time, where we determined bound
state solutions; in Sec. 4 we continue with the insertion of central
potentials in the mass term of the modified Klein-Gordon equation,
in which the potential considered is a Coulomb-type potential, in
which we define the relativistic energy profile of the quantum oscillator;
in Sec. 5, we present our conclusions.

\section{An overview of The Klein-Gordon oscillators in a cosmic string within
Kaluza-Klein Theory}

The purpose of this section is to investigate the dynamics of the
Klein-Gordon oscillator (KGO) in the framework of Kaluza-Klein theory
geometry. It is widely known that the relativistic wave equations
for a scalar particle in a Riemannian spacetime, described by the
metric tensor $g_{\mu\nu}$, can be obtained through a reformulation
of the Klein-Gordon (KG) equation. This provides an opportunity to
examine the behavior of the KGO in the presence of cosmic strings
\citep{key-B1, key-B2}. 
\begin{equation}
\left(\square+m^{2}-\mathcal{\xi}R\right)\Psi(x,t)=0,\label{eq:1}
\end{equation}
In this context, the symbol $\square$ denotes the Laplace-Beltrami
operator, which is commonly defined as follows: 
\begin{equation}
\square=g^{\mu\nu}D_{\mu}D_{\nu}=\frac{1}{\sqrt{-g}}\partial_{\mu}\left(\sqrt{-g}g^{\mu\nu}\partial_{\nu}\right),\label{eq:2}
\end{equation}
In the provided equation, the parameter $\xi$ corresponds to a dimensionless
coupling constant with real values. The Ricci scalar curvature, denoted
as $R$, can be expressed as $R=g^{\mu\nu}R_{\mu\nu}$, where $R_{\mu\nu}$
denotes the Ricci curvature tensor. The inverse of the metric tensor
is represented by $g^{\mu\nu}$, and the determinant of the metric
tensor is denoted as $g$, specifically $g=\det\left(g_{\mu\nu}\right)$.

Our subsequent aim is to investigate the quantum dynamics of spin-0
particles within the space-time generated by a $(4+1)$-dimensional
space. In this context, we will explore the behavior and properties
of these particles in this extended dimensional framework.

\subsection{Free Klein-Gordon equation in the background of a cosmic string in
a Kaluza--Klein theory}

In this section \citep{key-B3}, we embark on an exploration of a
fundamental topological defect that serves as the cornerstone of our
research. Taking inspiration from the intriguing characteristics observed
in edge dislocations present within crystalline solids, we extend
the concept of such defects into the realm of gravity. Upon examining
an edge dislocation, we discern its remarkable metamorphosis from
a circular form into a spiral dislocation. The line element that elucidates
the space-time curvature incorporating this significant topological
defect is expressed in units where $\hbar=c=1$ \citep{key-B3, key-B4}.
By establishing this connection, we bridge the principles governing
crystal defects with the intricate structure of space-time. 
\begin{align}
ds^{2} & =g_{\mu\nu}dx^{\mu}dx^{\nu}\nonumber \\
 & =dt^{2}-d\rho^{2}-\left(\alpha\rho\right)^{2}d\varphi^{2}-\left(dz+Jd\varphi\right)^{2}-\left(dx+\frac{\Phi}{2\pi}d\varphi\right)^{2},\label{eq:3}
\end{align}

In this context, where $\chi$ represents a constant value associated
with the distortion of the defect, the parameter $J=\frac{\left|\overrightarrow{b}\right|}{2\pi}$
also relates to the Burgers vector $\overrightarrow{b}$. The coordinate
ranges are as follows: $-\infty\le t\le+\infty$, $r\ge0$, $0\le\varphi\le2\pi$,
$-\infty\le z,x\le+\infty$, and $\alpha\in[0,1[$. The angular parameter
$\alpha$ determines the angular deficit $\delta\varphi=2\pi(1-\alpha)$,
which is connected to the linear mass density $\mu$ of the string
through $\alpha=1-4\mu$. It is important to note that this metric
provides an accurate solution to Einstein's field equations for $0\le\mu<1/4$.
Furthermore, by setting $\varphi^{\prime}=\alpha\varphi$, it represents
a flat conical outer space with an angular deficit $\delta\phi=8\pi\mu$.

When we consider the components of the metric tensor and its inverse,
denoted as $\left(g_{\mu\nu}\right)$ and $\left(g^{\mu\nu}\right)$,
respectively \citep{key-B3, key-B4}, we find the following expressions:
\begin{equation}
g_{\mu\nu}=\left(\begin{array}{ccccc}
1 & 0 & 0 & 0 & 0\\
0 & -1 & 0 & 0 & 0\\
0 & 0 & -\left(\left(\alpha\rho\right)^{2}+\left(\frac{\Phi}{2\pi}\right)^{2}+J^{2}\right) & -J & -\frac{\Phi}{2\pi}\\
0 & 0 & -J & -1 & 0\\
0 & 0 & -\frac{\Phi}{2\pi} & 0 & -1
\end{array}\right),\label{eq:4}
\end{equation}
and 
\begin{equation}
g^{\mu\nu}=\left(\begin{array}{ccccc}
1 & 0 & 0 & 0 & 0\\
0 & -1 & 0 & 0 & 0\\
0 & 0 & -\frac{1}{\left(\alpha\rho\right)^{2}} & \frac{J}{\left(\alpha\rho\right)^{2}} & \frac{\Phi}{2\pi\left(\alpha\rho\right)^{2}}\\
0 & 0 & \frac{J}{\left(\alpha\rho\right)^{2}} & -\left(1+\frac{J^{2}}{\left(\alpha\rho\right)^{2}}\right) & -\frac{\Phi J}{2\pi\left(\alpha\rho\right)^{2}}\\
0 & 0 & \frac{\Phi}{2\pi\left(\alpha\rho\right)^{2}} & -\frac{\Phi J}{2\pi\left(\alpha\rho\right)^{2}} & -\left(1+\frac{\Phi^{2}}{\left(2\pi\alpha\rho\right)^{2}}\right)
\end{array}\right)\label{eq:5}
\end{equation}
where 
\begin{equation}
\psi(t,\rho,\varphi,z,x)=\psi(\rho)e^{-i\left(Et-l\varphi-Kz-\lambda x\right)},\label{eq:6}
\end{equation}
\begin{equation}
\left[\frac{d^{2}}{d\rho^{2}}+\frac{1}{\rho}\frac{d}{d\rho}-\frac{\zeta^{2}}{\rho^{2}}+\kappa^{2}\right]\psi\left(\rho\right)=0,\label{eq:7}
\end{equation}
where we have set 
\begin{align}
\zeta= & \frac{\sqrt{\left(l-JK\right)^{2}+\frac{\Phi^{2}}{\pi^{2}}\left[\frac{\lambda^{2}}{4}-\frac{l\lambda}{\Phi}+\frac{JK\lambda\pi}{\Phi}\right]}}{\alpha}\label{eq:8}\\
\kappa= & \sqrt{E^{2}-m^{2}-K^{2}-\lambda^{2}}\nonumber 
\end{align}
The equation represented by Eq. \eqref{eq:7} can be identified as
a Bessel equation. Its general solution is defined as follows: 
\begin{equation}
\psi\left(\rho\right)=A\,J_{|\zeta|}\left(\kappa\rho\right)+B\,Y_{|\zeta|}\left(\kappa\rho\right),\label{eq:9}
\end{equation}
where $J_{|\zeta|}\left(\kappa\rho\right)$ and $Y_{|\zeta|}\left(\kappa\rho\right)$
are the Bessel functions of order $\zeta$ and of the first and the
second kind, respectively. Here $A$ and $B$ are arbitrary constants.
We notice that at the origin when $\zeta=0$, the function $J_{|\zeta|}\left(\kappa\rho\right)\ne0$.
However, $Y_{|\zeta|}\left(\kappa\rho\right)$ is always divergent
at the origin. In this case, we will consider only $J_{|\zeta|}\left(\kappa\rho\right)$
when $\zeta\ne0$. Hence, we write the solution to Eq. \eqref{eq:7}
as follows 
\begin{equation}
\psi\left(\rho\right)=A\,J_{\frac{\left|\sqrt{\left(l-JK\right)^{2}+\frac{\Phi^{2}}{\pi^{2}}\left[\frac{\lambda^{2}}{4}-\frac{l\lambda}{\Phi}+\frac{JK\lambda\pi}{\Phi}\right]}\right|}{\alpha}}\,\,\left(\sqrt{E^{2}-m^{2}-K^{2}-\lambda^{2}}\,\rho\right),\label{eq:10}
\end{equation}
We can now express the wavefunction of the spinless heavy KG particle
in the space-time of a cosmic dislocation using this solution. 
\begin{equation}
\psi\left(t,\rho,\varphi,z,x\right)=\left|\mathcal{N}_{1}\right|e^{-i\left(Et-l\varphi-Kz-\lambda x\right)}\,J_{\frac{\left|\sqrt{\left(l-JK\right)^{2}+\frac{\Phi^{2}}{\pi^{2}}\left[\frac{\lambda^{2}}{4}-\frac{l\lambda}{\Phi}+\frac{JK\lambda\pi}{\Phi}\right]}\right|}{\alpha}}\,\,\left(\sqrt{E^{2}-m^{2}-K^{2}-\lambda^{2}}\,\rho\right),\label{eq:11}
\end{equation}

The determination of the constant $\left|\mathcal{N}_{1}\right|$
can be achieved by applying the necessary normalization condition
to the Klein-Gordon equation. However, it is fortunate that the inability
to determine the normalization constants in this manuscript does not
impact the final outcomes or results.

\subsection{Klein-Gordon oscillator in the background of a cosmic string in a
Kaluza--Klein theory}

In order to proceed, it is necessary to substitute the momentum operator
into Equation \eqref{eq:1}. Consequently, Equation \eqref{eq:10}
can be reformulated as follows. Similarly, employing a straightforward
calculation based on the aforementioned approach, we can derive the
following differential equation \citep{key-B3}. 
\begin{equation}
\left[\frac{d^{2}}{d\rho^{2}}+\frac{1}{\rho}\frac{d}{d\rho}-m^{2}\omega^{2}\rho^{2}-\frac{\sigma^{2}}{\rho^{2}}+E^{2}+\delta\right]\psi\left(\rho\right)=0,\label{eq:12}
\end{equation}
with 
\begin{align}
\sigma^{2}= & \left(\frac{\left(l-JK\right)^{2}+\frac{\Phi^{2}}{\pi^{2}}\left[\frac{\lambda^{2}}{4}-\frac{l\lambda\pi}{\Phi}+\frac{JK\lambda\pi}{\Phi}\right]}{\alpha}\right)^{2}\label{eq:13}\\
\textcolor{black}{\delta-E^{2} }= & \textcolor{black}{ }2m\omega-m^{2}-K^{2}-\lambda^{2}\nonumber 
\end{align}

The Klein-Gordon oscillator (KGO) equation for a spin-0 particle within
the (1+4) space-time of Kaluza-Klein theory is represented by Equation
\eqref{eq:12}. In order to obtain the solution to this problem, we
propose a radial coordinate transformation as a preliminary step.
\begin{equation}
\mathcal{S}=m\omega r^{2},\label{eq:14}
\end{equation}
subsitutuing the expression for $\chi$ into Eq. \eqref{eq:12}, we
obtain 
\begin{equation}
\left[\frac{d^{2}}{d\mathcal{\mathcal{S}}^{2}}+\frac{1}{\mathcal{S}}\frac{\partial}{d\mathcal{\mathcal{S}}}-\frac{\sigma^{2}}{4\mathcal{\mathcal{S}}^{2}}+\frac{\delta}{4m\omega\mathcal{\mathcal{S}}}-\frac{1}{4}\right]\psi\left(\mathcal{S}\right)=0.\label{eq:15}
\end{equation}
By examining the asymptotic behavior of the wave function at both
the origin and infinity, and aiming to find regular solutions, we
can consider a solution of the following form: 
\begin{equation}
\psi\left(\mathcal{\mathcal{S}}\right)=\mathcal{\mathcal{S}}^{\frac{\left|\sigma\right|}{2}}e^{-\frac{\mathcal{\mathcal{S}}}{2}}F\left(\mathcal{\mathcal{S}}\right),\label{eq:16}
\end{equation}
As previously, we can plug this back into Eq. \eqref{eq:15}, and
we get 
\begin{equation}
\mathcal{S}\frac{d^{2}F\left(\mathcal{S}\right)}{d\mathcal{S}^{2}}+\left(|\sigma|+1-\chi\right)\frac{dF\left(\mathcal{S}\right)}{d\mathcal{S}}-\left(\frac{|\sigma|}{2}-\frac{\delta}{4m\omega}+\frac{1}{2}\right)F\left(\mathcal{S}\right)=0,\label{eq:17}
\end{equation}
The equation at hand corresponds to the confluent hypergeometric equation
\citep{key-64}, and its solutions are expressed in terms of a specific
type of confluent hypergeometric function. 
\begin{equation}
F\left(\mathcal{\mathcal{S}}\right)=_{1}F_{1}\left(\frac{\left|\sigma\right|}{2}-\frac{\delta}{4m\omega}+\frac{1}{2},+1,\mathcal{\mathcal{S}}\right),\label{eq:18}
\end{equation}

It is important to mention that the solution given by Equation \eqref{eq:18}
must be a polynomial function of degree $n$. However, as we let $n\rightarrow\infty$,
a divergence issue arises. In order to have a finite polynomial, the
factor associated with the last term in Equation \eqref{eq:17} must
be a negative integer. This condition implies that: 
\begin{equation}
\frac{\left|\sigma\right|}{2}-\frac{\delta}{4m\omega}+\frac{1}{2}=-n_{r}\qquad,n_{r}=0,1,2.....,\label{eq:19}
\end{equation}
Utilizing this result along with the parameters given in Equation
\eqref{eq:14}, we can derive the quantized energy spectrum of the
Klein-Gordon oscillator (KGO) in the space-time of a cosmic dislocation.
Consequently, we are able to determine the following: 
\begin{equation}
E^{\pm}\left(n_{r}\right)=\pm\sqrt{4m\omega n_{r}+\frac{2m\omega}{\alpha}\left|\left(l-JK\right)^{2}+\frac{\Phi^{2}}{\pi^{2}}\left[\frac{\lambda^{2}}{4}-\frac{l\pi\lambda}{\Phi}+\frac{JK\lambda\pi}{\Phi}\right]\right|+m^{2}+K^{2}+\lambda^{2}},\label{eq:20}
\end{equation}
It is evident that the energy is directly dependent on the angular
deficit $\alpha$. In other words, the presence of the wedge angle,
which is caused by the topological defect (i.e., the cosmic string),
influences the curvature of the space-time. As a result, it affects
the relativistic dynamics of the scalar particle by introducing a
gravitational field. The corresponding wave function is defined as
follows: 
\begin{equation}
\psi\left(\rho\right)=\left|\mathcal{N}_{2}\right|\left(m\omega\rho^{2}\right)^{\frac{\left|\sigma\right|}{2}}e^{-\frac{m\omega\rho^{2}}{2}}{}_{1}F_{1}\left(\frac{\left|\sigma\right|}{2}-\frac{\delta}{4m\omega}+\frac{1}{2},\sigma+1,m\omega\rho^{2}\right),\label{eq:21}
\end{equation}
Subsequently, the general eigenfunctions are expressed as: 
\begin{equation}
\psi\left(t,\rho,\varphi,z,x\right)=\left|\mathcal{N}_{2}\right|\left(m\omega\rho^{2}\right)^{\frac{\left|\sigma\right|}{2}}e^{-\frac{m\omega\sigma^{2}}{2}}e^{-i\left(Et-l\varphi-Kz-\lambda x\right)}{}_{1}F_{1}\left(\frac{\left|\sigma\right|}{2}-\frac{\delta}{4m\omega}+\frac{1}{2},\sigma+1,m\omega\rho^{2}\right),\label{eq:22}
\end{equation}
Here, $\left|\mathcal{N}_{2}\right|$ represents the normalization
constant, and the values of $\left(\sigma,\delta\right)$ are provided
in Equation \eqref{eq:13}.

\section{PDM of Klein-Gordon oscillator in the background of a cosmic string
in a Kaluza-Klein theory}

In this subsection, we shall use the non-minimal coupling form of
the PDM-momentum operator
\begin{equation}
\mathbf{\hat{p}}\left(\mathbf{r}\right)=-i\left(\mathbf{\nabla-}\frac{\mathbf{\nabla}f\left(\mathbf{r}\right)}{4f\left(\mathbf{r}\right)}\right)\Longleftrightarrow p_{j}=-i\left(\partial_{j}-\frac{\partial_{j}f\left(\mathbf{r}\right)}{4f\left(\mathbf{r}\right)}\right),\label{Om1}
\end{equation}
introduced by Mustafa and Algadhi \cite{O1}. It has been shown that
this PDM-momentum operator effectively yields the what is known in
the literature as Mustafa and Mazharimousavi's ordering \cite{O2,O3}
of the von Roos PDM-kinetic energy Schrödinger operator \cite{O4}.
It has been used in the study of PDM Klein-Gordon (KG) particles in
different spacetime backgrounds \cite{O5,O6,O7,O8,O9}. Under such
non-minimal coupling form, our PDM operator reads

\begin{equation}
\tilde{p}_{\mu}\longrightarrow-i\partial_{\mu}+i\mathcal{F}_{\mu}\,;\,\,\mathcal{F}_{\mu}=\left(0,\mathcal{F}_{\rho},0,0\right),\,\,\mathcal{F}_{\rho}=\frac{f^{\prime}\left(\rho\right)}{4f\left(\rho\right)}\label{Om2}
\end{equation}
and consequently our PDM KG-equation reads
\begin{equation}
\frac{1}{\sqrt{-g}}\left(D_{\mu}+\mathcal{F}_{\mu}\right)\left[\sqrt{-g}g^{\mu\nu}\left(D_{\nu}-\mathcal{F}_{\nu}\right)\Psi\right]=\left(m+S\left(\rho\right)\right)^{2}\Psi.\label{Om3}
\end{equation}
where $D_{\mu}=\partial_{\mu}-ieA_{\mu}$ is the covariant derivative
and $S\left(\rho\right)$ is the Lorentz scalar radial potential.
This would, with $\Psi$ given in (\ref{eq:11}), imply that
\begin{equation}
\psi^{\prime\prime}\left(\rho\right)+\frac{1}{\rho}\psi^{\prime}\left(\rho\right)+\left[\mathcal{E}^{2}-\frac{\tilde{\zeta}^{2}}{\rho^{2}}-\mathcal{M}\left(\rho\right)-2mS\left(\rho\right)-S\left(\rho\right)^{2}\right]\psi\left(\rho\right)=0,\label{Om4}
\end{equation}
where $\mathcal{E}^{2}=E^{2}-\left(K^{2}+\lambda^{2}+m^{2}\right)$,
and 
\begin{equation}
\tilde{\zeta}^{2}=\frac{1}{\alpha^{2}}\left\{ \left(\left[\ell-eA_{\varphi}\right]-JK\right)^{2}+\lambda\,\bar{\phi}\left[\lambda\,\bar{\phi}-2\left(\ell-eA_{\varphi}\right)+2JK\right]\right\} ;\bar{\phi}=\frac{\phi}{2\pi},\label{Om5}
\end{equation}
and
\begin{equation}
\mathcal{M}\left(\rho\right)=\mathcal{F}_{\rho}^{\prime}+\frac{\mathcal{F}_{\rho}}{\rho}+\mathcal{F}_{\rho}^{2}.\label{Om6}
\end{equation}
The substitution of $\psi\left(\rho\right)=U\left(\rho\right)/\sqrt{\rho}$
would imply
\begin{equation}
U^{\prime\prime}\left(\rho\right)+\left[\mathcal{E}^{2}-\frac{\left(\tilde{\zeta}^{2}-1/4\right)}{\rho^{2}}-\mathcal{M}\left(\rho\right)-2mS\left(\rho\right)-S\left(\rho\right)^{2}\right]U\left(\rho\right)=0.\label{Om7}
\end{equation}
Eventually, with $\mathcal{F}_{\rho}=\eta\rho$ (or if you wish, $f\left(\rho\right)=\exp\left(2\eta\rho\right)$)
of (\ref{Om2}), we obtain
\begin{equation}
U^{\prime\prime}\left(\rho\right)+\left[\mathcal{\tilde{E}}^{2}-\frac{\left(\tilde{\zeta}^{2}-1/4\right)}{\rho^{2}}-\mathcal{\eta}^{2}\rho^{2}-2mS\left(\rho\right)-S\left(\rho\right)^{2}\right]U\left(\rho\right)=0.\label{Om8}
\end{equation}
where $\mathcal{\tilde{E}}^{2}=\mathcal{E}^{2}-2\eta$. At this point,
one may observe that for $\eta=m\omega$, $A_{\varphi}=0$, and $S\left(\rho\right)=0$,
the case discussed for (\ref{eq:12}) is retrieved to obtain
\begin{equation}
\mathcal{\tilde{E}}^{2}=2\eta\left(2n_{r}+\left\vert \tilde{\zeta}\right\vert +1\right)\Leftrightarrow E=\pm\sqrt{2\eta\left(2n_{r}+\left\vert \tilde{\zeta}\right\vert +2\right)+K^{2}+\lambda^{2}+m^{2}},\label{Om9}
\end{equation}
and
\begin{equation}
U\left(\rho\right)\sim\rho^{\left\vert \tilde{\zeta}\right\vert +1/2}\exp\left(-\frac{\eta\rho^{2}}{2}\right)L_{n_{r}}^{\left\vert \tilde{\zeta}\right\vert }\left(\left\vert \tilde{\zeta}\right\vert \rho^{2}\right).\label{Om10}
\end{equation}

\subsection{\textcolor{black}{KG-oscillators plus linear confinement $S\left(\rho\right)=A\rho$}}

With a linear Lorentz scalar potential $S\left(\rho\right)=A\rho$,
equation (\ref{Om8}) reads
\begin{equation}
U^{\prime\prime}\left(\rho\right)+\left[\mathcal{\tilde{E}}^{2}-\frac{\left(\tilde{\zeta}^{2}-1/4\right)}{\rho^{2}}-\mathcal{\tilde{\eta}}^{2}\rho^{2}-2mA\rho\right]U\left(\rho\right)=0\;;\;\mathcal{\tilde{\eta}}^{2}=\mathcal{\eta}^{2}+A^{2}.\label{Om11}
\end{equation}
Hereby, we shall be interested in an exact solution for equation (\ref{Om11})
for $\tilde{\zeta}=1/2\Leftrightarrow4\left(\ell-JK-\lambda\bar{\phi}\right)^{2}=\alpha^{2}$,
so that equation (\ref{Om11}) reads 
\begin{equation}
\left\{ \partial_{\rho}^{2}-\mathcal{\tilde{\eta}}^{2}\rho^{2}-2mA\,\rho+\mathcal{\tilde{E}}^{2}\right\} U\left(\rho\right)=0.\label{Om12}
\end{equation}
At this point, one should be aware that this equation resembles the
radially spherically symmetric Schrödinger equation with an irrational
angular momentum quantum number $\tilde{\ell}=\tilde{\zeta}-1/2=0$
in the central attractive/repulsive core $\tilde{\ell}(\tilde{\ell}+1)/\rho^{2}$.
Let us use a radial wave function in the form of
\begin{equation}
U\left(\rho\right)=\exp\left(-\frac{|\tilde{\eta}|\rho^{2}}{2}-\frac{mA\rho}{|\tilde{\eta}|}\right)\,\left[B_{1}\,(\rho+\frac{mA}{\mathcal{\tilde{\eta}}^{2}})\,F(\rho)+B_{2}\,G(\rho)\right];\,\,\tilde{A}=\frac{mA}{\mathcal{\tilde{\eta}}^{2}}.\label{Om13}
\end{equation}
Obviously, the exponential term in (\ref{Om13}) represents the asymptotic
behaviour $U(\rho)\rightarrow0$ as $\rho\rightarrow\infty$. Whereas,
the asymptotic behaviour of $U(\rho)$ as $\rho\rightarrow0$ is either
finite or zero, depending on the effective interaction potential at
hand ( e.g., \cite{O10}). We shall instead let the general functions
$F(\rho)$ and $G(\rho)$ have their say in the process. Hence, the
substitution of $U(\rho)$, (\ref{Om13}), would in a straightforward
manner result in
\begin{eqnarray}
 &  & B_{1}\,\left[(\rho+\tilde{A})\,\partial_{\rho}^{2}-2|\tilde{\eta}|\left((\rho+\tilde{A})^{2}-\frac{1}{|\tilde{\eta}|}\right)\,\partial_{\rho}+[\mathcal{\tilde{E}}^{2}-3|\tilde{\eta}|+\frac{m^{2}A^{2}}{\tilde{\eta}^{2}}]\right]F(\rho)\nonumber \\
 &  & +\,B_{2}\,\left[(\rho+\tilde{A})\,\partial_{\rho}^{2}-2|\tilde{\eta}|((\rho+\tilde{A})\,\partial_{\rho}+[\mathcal{\tilde{E}}^{2}-|\tilde{\eta}|+\frac{m^{2}A^{2}}{\tilde{\eta}^{2}}]\right]G(\rho)=0.\label{Om14}
\end{eqnarray}
This equation suggests that
\begin{equation}
\left[(\rho+\tilde{A})\,\partial_{\rho}^{2}-2|\tilde{\eta}|\left((\rho+\tilde{A})^{2}-\frac{1}{|\tilde{\eta}|}\right)\,\partial_{\rho}+[\mathcal{\tilde{E}}^{2}-3|\tilde{\eta}|+\frac{m^{2}A^{2}}{\tilde{\eta}^{2}}]\right]F(\rho)=0,\label{Om15}
\end{equation}
and
\begin{equation}
\left[(\rho+\tilde{A})\,\partial_{\rho}^{2}-2|\tilde{\eta}|((\rho+\tilde{A})\,\partial_{\rho}+[\mathcal{\tilde{E}}^{2}-|\tilde{\eta}|+\frac{m^{2}A^{2}}{\tilde{\eta}^{2}}]\right]G(\rho)=0,\label{Om16}
\end{equation}
A change of variable a change of variable $y=|\mathcal{\tilde{\eta}}|(\rho+\tilde{A})^{2}$
would imply
\begin{equation}
\left[y\,\partial_{y}^{2}+(\frac{3}{2}-y)\,\partial_{y}+\frac{1}{4|\mathcal{\tilde{\eta}}|}[\mathcal{\tilde{E}}^{2}-3|\mathcal{\tilde{\eta}}|+\frac{m^{2}A^{2}}{\mathcal{\tilde{\eta}}^{2}}]\right]\,F(y)=0,\label{Om17}
\end{equation}
and
\begin{equation}
\left[y\,\partial_{y}^{2}+(\frac{1}{2}-y)\,\partial_{y}+\frac{1}{4|\mathcal{\tilde{\eta}}|}[\mathcal{\tilde{E}}^{2}-|\mathcal{\tilde{\eta}}|+\frac{m^{2}A^{2}}{\mathcal{\tilde{\eta}}^{2}}]\right]\,G(y)=0.\label{Om18}
\end{equation}
One should be able to observe that both equations are in the form
of Kummer's equation
\begin{equation}
\lbrack u\,\partial_{u}^{2}+(b-u)\,\partial_{u}-a]W(u)=0.\label{Om19}
\end{equation}
to immediately imply that 
\begin{equation}
F(y)=\,_{1}F_{1}(-\frac{1}{4|\mathcal{\tilde{\eta}}|}(\mathcal{\tilde{E}}^{2}-3|\mathcal{\tilde{\eta}}|+\frac{m^{2}A^{2}}{\mathcal{\tilde{\eta}}^{2}}),\frac{3}{2},y),\label{Om20}
\end{equation}
and
\begin{equation}
G(\eta)=\,_{1}F_{1}(-\frac{1}{4|\mathcal{\tilde{\eta}}|}(\mathcal{\tilde{E}}^{2}-|\mathcal{\tilde{\eta}}|+\frac{m^{2}A^{2}}{\mathcal{\tilde{\eta}}^{2}}),\frac{1}{2},y),\label{Om21}
\end{equation}
Our general solution for (\ref{Om13}) therefore reads
\begin{eqnarray}
U\left(\rho\right) & = & \exp\left(-\frac{|\tilde{\eta}|\rho^{2}}{2}-\frac{mA\rho}{|\tilde{\eta}|}\right)\,\left[B_{1}\,(\rho+\frac{mA}{\mathcal{\tilde{\eta}}^{2}})\,_{1}F_{1}(-\frac{1}{4|\mathcal{\tilde{\eta}}|}(\mathcal{\tilde{E}}^{2}-3|\mathcal{\tilde{\eta}}|+\frac{m^{2}A^{2}}{\mathcal{\tilde{\eta}}^{2}}),\frac{3}{2},y)\right.\nonumber \\
 &  & \text{ \ \ \ \ \ \ \ \ \ \ \ \ \ \ \ \ \ \ \ \ \ \ \ \ \ \ \ \ \ \ \ \ \ \ \ \ \ }\left.+B_{2}\,\,_{1}F_{1}(-\frac{1}{4|\mathcal{\tilde{\eta}}|}(\mathcal{\tilde{E}}^{2}-|\mathcal{\tilde{\eta}}|+\frac{m^{2}A^{2}}{\mathcal{\tilde{\eta}}^{2}}),\frac{1}{2},y)\right]\label{Om22}
\end{eqnarray}
Now we have to appeal to textbook exact solvability of a pure harmonic
oscillator $\mathcal{\tilde{\eta}}^{2}\rho^{2}$, with $A=0$, and
enforce the condition that $U\left(\rho\right)=0$ for $\rho\rightarrow0$.
Consequently, $B_{2}=0$ and our sought-after solution is obviously
given, with $y=|\mathcal{\tilde{\eta}}|(\rho+\tilde{A})^{2}$, by
\begin{equation}
U\left(\rho\right)=B_{1}\,(\rho+\frac{mA}{\mathcal{\tilde{\eta}}^{2}})\,\exp\left(-\frac{|\tilde{\eta}|\rho^{2}}{2}-\frac{mA\rho}{|\tilde{\eta}|}\right)\,\,_{1}F_{1}(-\frac{1}{4|\mathcal{\tilde{\eta}}|}(\mathcal{\tilde{E}}^{2}-3|\mathcal{\tilde{\eta}}|+\frac{m^{2}A^{2}}{\mathcal{\tilde{\eta}}^{2}}),\frac{3}{2},y).\label{Om23}
\end{equation}
However, finiteness and square integrability of the quantum mechanics
wave functions requires that the confluent hypergeometric series should
be truncated into a polynomial of order $n_{r}=0,1,2,\cdots$. Under
such condition
\begin{equation}
-n_{r}=-\frac{1}{4|\mathcal{\tilde{\eta}}|}(\mathcal{\tilde{E}}^{2}-3|\mathcal{\tilde{\eta}}|+\frac{m^{2}A^{2}}{\mathcal{\tilde{\eta}}^{2}})\Longrightarrow\mathcal{\tilde{E}}^{2}=\mathcal{E}^{2}-2\left\vert \eta\right\vert =2|\mathcal{\tilde{\eta}}|\left(2n_{r}+\frac{3}{2}\right)-\frac{m^{2}A^{2}}{\mathcal{\tilde{\eta}}^{2}},\label{Om24}
\end{equation}
and 
\begin{equation}
E=\pm\sqrt{2|\mathcal{\tilde{\eta}}|\left(2n_{r}+\frac{3}{2}\right)+2\left\vert \eta\right\vert +\left(K^{2}+\lambda^{2}+m^{2}\right)-\frac{m^{2}A^{2}}{\mathcal{\tilde{\eta}}^{2}}}.\label{Om25}
\end{equation}
It should be noted that this results is in exact accord with that
in (\ref{Om9}) for $\tilde{\zeta}=1/2$ and $A=0$. Our solution
in (\ref{Om23}) suggests, beyond doubt, that a quantum particle moving
in an oscillator-plus-linear radial potential is allowed to reach
the center at $\rho=0$ with some amplitude ( e.g., \cite{O10}).
Then the conjecture that $U(\rho)\rightarrow0$ as $\rho\rightarrow0$
is not a general admissible condition in quantum mechanics, but finiteness
and square integrability of the radial wave function is the only valid
condition in general.

On the other hand, it could be interesting to know that equation (\ref{Om11})
admits a solution in the form of biconfluent Heun functions
\begin{equation}
U\left(\rho\right)=C_{1}\,\rho^{\left\vert \tilde{\zeta}\right\vert +1/2}\exp\left(-\frac{|\tilde{\eta}|\rho^{2}}{2}-\frac{mA\rho}{|\tilde{\eta}|}\right)H_{B}\left(\alpha^{\prime},\beta^{\prime},\gamma^{\prime},\delta^{\prime},\sqrt{|\tilde{\eta}|}\rho\right),\label{Om26}
\end{equation}
where
\begin{equation}
\alpha^{\prime}=2\left\vert \tilde{\zeta}\right\vert ,\;\beta^{\prime}=\frac{2mA}{|\tilde{\eta}|^{3/2}},\;\gamma^{\prime}=\frac{m^{2}A^{2}+\mathcal{\tilde{E}}^{2}\tilde{\eta}^{2}}{\tilde{\eta}|^{3}},\;\delta^{\prime}=0.\label{Om27}
\end{equation}
This would result, for $\tilde{\zeta}=1/2$,
\begin{equation}
U\left(\rho\right)=C_{1}\,\rho\,\exp\left(-\frac{|\tilde{\eta}|\rho^{2}}{2}-\frac{mA\rho}{|\tilde{\eta}|}\right)H_{B}\left(1,\beta^{\prime},\gamma^{\prime},0,\sqrt{|\tilde{\eta}|}\rho\right).\label{Om28}
\end{equation}
A simple comparison between (\ref{Om23}) and (\ref{Om28}) would
suggest, up to a constant, that
\begin{equation}
(\rho+\frac{mA}{\mathcal{\tilde{\eta}}^{2}})\,\,\,_{1}F_{1}(-\frac{1}{4|\mathcal{\tilde{\eta}}|}(\mathcal{\tilde{E}}^{2}-3|\mathcal{\tilde{\eta}}|+\frac{m^{2}A^{2}}{\mathcal{\tilde{\eta}}^{2}}),\frac{3}{2},|\mathcal{\tilde{\eta}}|(\rho+\tilde{A})^{2})=\rho\,\,H_{B}\left(1,\beta^{\prime},\gamma^{\prime},0,\sqrt{|\tilde{\eta}|}\rho\right)\label{Om29}
\end{equation}
to imply
\begin{equation}
(x+\frac{\beta^{\prime}}{2})\,_{1}F_{1}\left(\frac{3-\gamma^{\prime}}{4},\,\frac{3}{2},(x+\frac{\beta^{\prime}}{2})^{2}\right)=x\,H_{B}(1,\beta^{\prime},\gamma^{\prime},0,x);\,\,x=\sqrt{|\tilde{\eta}|}\rho.\label{Om30}
\end{equation}
Hence, the truncation of the bi-confluent Heun series $H_{B}\left(\alpha^{\prime},\beta^{\prime},\gamma^{\prime},\delta^{\prime},\sqrt{|\tilde{\eta}|}\rho\right)$
into a polynomial of of degree $n\geq0$, , one has to apply the condition
that $\gamma^{\prime}=2\left(n+1\right)+\alpha^{\prime}\Longrightarrow\gamma^{\prime}=2\left(n+3/2\right)$
for $\alpha^{\prime}=1$. Whereas the truncation of the confluent
geometric series into a polynomial of order $n_{r}=0,1,2,\cdots$
suggests that\ $-4n_{r}=3-\gamma^{\prime}\Longrightarrow\gamma^{\prime}=2(2n_{r}+3/2$.
Consequently, the confluent geometric and bi-confluent Heun polynomials
are both of even powers and $n=2n_{r}\geq0$. Moreover, our result
in (\ref{Om30}) would, for $\beta^{\prime}=0$, yield the correlation
\begin{equation}
\,_{1}F_{1}\left(\frac{3-\gamma^{\prime}}{4},\,\frac{3}{2},x^{2}\right)=\,H_{B}(1,0,\gamma^{\prime},0,x),\label{Om31}
\end{equation}
which is, in fact, in exact accord with that reported by Ronveaux
\cite{O11} 
\begin{equation}
H_{B}(\alpha^{\prime},0,\gamma^{\prime},0,y)=\,_{1}F_{1}\left(\frac{1}{2}+\frac{\alpha^{\prime}}{4}-\frac{\gamma^{\prime}}{4},1+\frac{\alpha^{\prime}}{2},y^{2}\right).\label{Om32}
\end{equation}

\section{ Free KG equation in the background of a cosmic string in a Kaluza--Klein
theory under coulomb-type potential}

In this section, we are studying the interaction between the K-G equation
and Coulomb-type potential in KKT, in the space-time caused by a (4+1)-dimensional
stationary cosmic string described in Coulomb-Type Potentials $\left(S\left(r\right)=\frac{\kappa}{r}=\pm\frac{\left|\kappa\right|}{r}\right)$\citep{key-012}.
\begin{equation}
\left(\frac{d^{2}}{d\rho^{2}}+\frac{1}{\rho}\frac{d}{d\rho}-\frac{\sigma^{2}}{\rho^{2}}+\left(E-S\left(\rho\right)\right)^{2}-\xi\right)\psi\left(\rho\right)=0
\end{equation}
where 
\begin{equation}
\left(\frac{d^{2}}{d\rho^{2}}+\frac{1}{\rho}\frac{d}{d\rho}-\frac{\sigma^{2}}{\rho^{2}}+\left(E-\frac{\kappa}{\rho}\right)^{2}-\xi\right)\psi\left(\rho\right)=0
\end{equation}
and 
\begin{equation}
\xi=\left(m^{2}+K^{2}+\lambda^{2}\right)-\left(\frac{\Omega}{\alpha}\right)^{2}\left(\ell-JK-\frac{\Phi}{2\pi}\lambda\right)^{2}
\end{equation}
By simplifying the last differential equation we find 
\begin{equation}
\left(\frac{d^{2}}{d\rho^{2}}+\frac{1}{\rho}\frac{d}{d\rho}-\frac{\left(\sigma^{2}-\kappa^{2}\right)}{\rho^{2}}-\frac{2\kappa E}{\rho}+E^{2}-\xi\right)\psi\left(\rho\right)=0
\end{equation}
The solution to this differential equation is the following: 
\begin{equation}
\psi\left(\rho\right)=\mathcal{N}\rho^{\alpha}\mathrm{WhittakerM}\left[\beta,\gamma,\vartheta\,\rho\right]
\end{equation}
With derivation and compensation, we find the following constants
\begin{equation}
\begin{array}{c}
\alpha=-1/2\\
\beta=-\frac{\kappa E}{\sqrt{\xi-E^{2}}}\\
\gamma=\mathrm{i}\left(\sqrt{\kappa^{2}-\sigma^{2}}\right)\\
\vartheta=2\sqrt{\xi-E^{2}}
\end{array}
\end{equation}
The general solution of the last differential equation is given in
the following form: 
\begin{equation}
\psi\left(\rho\right)=\mathcal{N}\rho^{-\frac{1}{2}}\mathrm{WhittakerM}\left[-\frac{\kappa E}{\sqrt{\xi-E^{2}}},\mathrm{i}\left(\sqrt{\kappa^{2}-\sigma^{2}}\right),2\sqrt{\xi-E^{2}}\,\rho\right]
\end{equation}
The mathematical relationship between the WhitakarM function and the
hypergeometric function ($1F1$) is given from the general form\citep{key-012,key-013}
\begin{equation}
\mathrm{WhittakerM}\left(\mu,\nu,z\right)=\mathrm{e}^{-\frac{z}{2}}z^{\frac{1}{2}+\nu}\mathrm{hypergeom}\left(\left[\frac{1}{2}+\nu-\mu\right],\left[1+2\nu\right],z\right)
\end{equation}
where the wave function of the free KG equation in this system given
as 
\begin{equation}
\psi\left(\rho\right)=\mathcal{N}\left(\mathit{\rho^{-\frac{1}{2}}\,\mathrm{e}^{-\left(\sqrt{\xi-E^{2}}\right)\,\rho}}\left(2\left(\sqrt{\xi-E^{2}}\right)\,\rho\right){}^{\frac{1}{2}+\mathrm{i}\left(\sqrt{\kappa^{2}-\sigma^{2}}\right)}\right)R\left(\rho\right)
\end{equation}
and 
\begin{equation}
R\left(\rho\right)={}_{1}F_{1}\left(\left[\frac{2\,\mathrm{i}\left(\sqrt{\kappa^{2}-\sigma^{2}}\right)\,\left(\sqrt{\kappa-E^{2}}\right)+2E\kappa}{2\left(\sqrt{\xi-E^{2}}\right)}+\frac{1}{2}\right],\left[1+2\,\mathrm{i}\left(\sqrt{\kappa^{2}-\sigma^{2}}\right)\right],2\left(\sqrt{\xi-E^{2}}\right)\,\rho\right)
\end{equation}
using the condition of quantification 
\begin{equation}
\frac{2\,\mathrm{i}\left(\sqrt{\kappa^{2}-\sigma^{2}}\right)\,\left(\sqrt{\kappa-E^{2}}\right)+2E\kappa}{2\left(\sqrt{\xi-E^{2}}\right)}+\frac{1}{2}=-n
\end{equation}
After the mathematical simplification of the equation, We find the
energy phrase given in the following form: 
\begin{equation}
\begin{array}{c}
E^{\pm}\left(n\right)=\pm\left(\frac{16\left(\sqrt{\xi^{2}\kappa^{2}\left(2n+1\right){}^{2}\left(\kappa^{2}-\sigma^{2}\right)}\right)}{16\left[n^{4}+2n^{3}+\sigma^{4}+4\kappa^{4}\right]+\left[32\sigma^{2}+24\right]n^{2}+\left[32\sigma^{2}+8\right]n+8\left[-8\kappa^{2}+1\right]\sigma^{2}+1}\right)\\
\pm\left(\frac{\xi\left[16\left(n^{4}+2n^{3}\right)+\left(-16\left[\kappa^{2}+2\sigma^{2}\right]+24\right)n^{2}+\left(-16\left[\kappa^{2}+2\sigma^{2}\right]+8\right)n+16\sigma^{4}+8\left(-6\kappa^{2}+1\right)\sigma^{2}+4\left[8\kappa^{4}-\kappa^{2}\right]+1\right]}{16\left[n^{4}+2n^{3}+\sigma^{4}+4\kappa^{4}\right]+\left[32\sigma^{2}+24\right]n^{2}+\left[32\sigma^{2}+8\right]n+8\left[-8\kappa^{2}+1\right]\sigma^{2}+1}\right)
\end{array}
\end{equation}
This last equation represents the energy of Klein-Gordon's free equation
as a quantitative number $n$ where $E^{\pm}\left(n\right)=f(n)$.
We note as a result that this last energy is a function dependent
of $n$ compared to the result in the first part by the connotation
of the Bessel function, which explains the actual effect of Coulomb-type
potential.

\section{Conclusion }

We have been investigating KGO immersed in a non-trivial background
characterized by ingredients that characterize the presence of curvature,
torsion and extra dimension, that is, a KKT spacetime with cosmic
string and screw-like dislocation, associated with curvature and torsion,
respectively. Furthermore, the extra dimension of KKT provides a quantum
flux, which influences the quantum dynamics of KGO.

Through a purely analytical analysis, we show that the KGO is influenced
not only by curvature, torsion and the quantum flux associated with
the extra dimension, but also by the possibility of modifying the
mass term of the Klein-Gordon equation.

The influence of curvature and torsion on the KGO can be visualized
through a redefinition of the angular momentum eigenvalues, which
are now dependent on the parameters associated with the cosmic string
and the screw-like displacement. Furthermore, the quantum flux coming
from the extra dimension is part of this quantum effect of "correcting"
the angular momentum eigenvalues. These quantum effects can be seen
as an effect analogous to ABE.

The modification of the mass term of the Klein-Gordon equation drastically
modifies the energy profile of the KGO, that is, the central potentials
inserted in the axial equation provide totally different relativistic
energy profiles.

....................................................


\begin{thebibliography}{10}
\bibitem{bd} N. D. Birrel, P. C. W. Davies, \textit{{Quantum fields
in curved space}}, 1nd edn. (Cambrigde University Press, Cambridge,
1984).

\bibitem{ryder} L. H. Ryder, \textit{{Quantum Field Theory}}, 2nd edn. (Cambrigde University Press, Cambridge, 1996).

\bibitem{kleinert} H. Kleinert, \textit{{Gauge fields in condensed matter}}, Vol2, (World Scientific, Singapore, 1989).

\bibitem{bas} G. Bastard, \textit{{Wave Mechanics Applied to Semiconductor
Heterostructure}}, 1nd edn. (Les Editions de Physique, Les Ulis 1988).

\bibitem{dantas} L. Dantas, C. Furtado, A. L. Silva Netto, Phys. Lett. A \textbf{{379}}, 11 (2015).

\bibitem{graff} M. J. Bueno, C. Furtado, A. M. de M. Carvalho, Eur.
Phys. J. B \textbf{{85}}, 53 (2012).

\bibitem{graff1} A. M. de M. Carvalho, C. A. de Lima Ribeiro, F.
Moraes, C. Furtado, Eur. Phys. J. Plus \textbf{{128}}, 60 (2013).

\bibitem{kat} M. O. Katanaev, I.V. Volovich, Ann. Phys. \textbf{{216}},
1 (1992).

\bibitem{kat1} K. C. Valanis, V.P. Panoskaltsis, Acta Mech. \textbf{{175}},
77 (2005).

\bibitem{eug} E. R. Figueiredo Medeiros, E. R. Bezerra de Mello,
Eur. Phys. J. C \textbf{{72}}, 2051 (2012).

\bibitem{bou} A. Boumali and N. Messai, Can. J. Phys. \textbf{{92}},
(2014).

\bibitem{vil} A. Vilenkin, E.P.S. Shellard, \textit{{Strings and
Other Topological Defects}} (Cambrigde University Press, Cambridge,
1994).

\bibitem{sc} T. W. B. Kibble, J. Phys. A: Math. Gen. \textbf{{9}},
1387 (1976).

\bibitem{put} R. A. Puntigam, H. H. Soleng, Class. Quantum Grav.
\textbf{{14}}, 1129 (1997).

\bibitem{ric} R. L. L. Vitória, K. Bakke, Gen. Relativ. Gravit. \textbf{{48}},
161 (2016).

\bibitem{ric1} R. L. L. Vitória, K. Bakke, Eur. Phys. J. Plus \textbf{{133}},
490 (2018).

\bibitem{ric2} R. L. L. Vitória, K. Bakke, Int. J. Mod. Phys. D \textbf{{27}},
1850005 (2018).

\bibitem{ric3} R. L. L. Vitória, K. Bakke, Eur. Phys. J. C \textbf{{78}},
175 (2018).

\bibitem{ric4} K. Bakke, V. B. Bezerra, R. L. L. Vitória, Int. J.
Mod. Phys. A \textbf{{35}}, 2050129 (2020).

\bibitem{abe} Y. Aharonov, D. Bohm, Phys. Rev. \textbf{{115}},
485 (1959).

\bibitem{abe1} M. Peshkin, A. Tonomura, \textit{{The Aharonov--Bohm
Effect in Lecture Notes in Physics}}, vol. \textbf{{340}} (Springer,
Berlin, 1989).

\bibitem{valdir} V. B. Bezerra, Phys. Rev. D \textbf{{35}}, 2031
(1987).

\bibitem{valdir1} V. B. Bezerra, Phys. Rev. D \textbf{{36}}, 1936
(1988).

\bibitem{erico} E. V. B. Leite, H. Belich, K. Bakke, Adv. High Energy
Phys. \textbf{{2015}}, 1 (2015).

\bibitem{kkt} Th. Kaluza, Sitzungsber. K. Preuss. Akad. Wiss. \textbf{{K1}},
966 (1921).

\bibitem{kkt1} O. Z. Klein, Phys. Z. \textbf{{37}}, 895 (1926).

\bibitem{kkt2} M. B. Green, J. H. Schwarz and E. Witten, \textit{{Superstring
Theory}}, Vols. 1, 2 (Cambridge Univ. Press, 1987).

\bibitem{erico1} E. V. B. Leite, R. L. L. Vitória, H. Belich, Modern
Physics Letters A \textbf{{34}}, 1950319 (2019).

\bibitem{erico2} E. V. B. Leite, H. Belich, R. L. L. Vitória, Adv.
High Energy Phys. \textbf{{2019}}, 6740360 (2019)

\bibitem{erico3} E. V. B. Leite, H. Belich, R. L. L. Vitória, Braz.
J. Phys. \textbf{{50}}, 744 (2020).

\bibitem{erico4} E. V. B. Leite, H. Belich, R. L. L. Vitória, Mod.
Phys. Lett. A \textbf{{35}}, 2050283 (2020).

\bibitem{boumali1}A. Boumali. EJTP \textbf{32}, 121-130 (2015).

\bibitem{boumali2}A. Boumali and L. chetouani. Phys. Lett. A \textbf{346},
261-268 (2005).

\bibitem{boumali3}A. Boumali and H. Aounallah, Advances in High Energy
Physics \textbf{2018} (2018).

\bibitem{boumali4}H. Aounallah, A. Boumali. Phys. Part. Nuclei Lett.
\textbf{16}, 195-205 (2019).

\bibitem{boumali5}A. Boumali and H. Aounallah, Rev. Mex. Fis. \textbf{66},
192-208 (2020).

\bibitem{key-B1} F. L. Gross, Relativistic quantum mechanics and
field theory, 1993.

\bibitem{key-B2} O. Klein, Z. Phys \textbf{37}, 895 (1926).

\bibitem{key-B3} A. Bouzenada, A. Boumali, R. L. L. Vitoria, F. Ahmed,
and M. Al-Raeei, (2023).arXiv preprint arXiv:2305.00793.

\bibitem{key-B4} J. Carvalho, et al, Eur. Phys. J. C 76, 1-9 (2016).

\bibitem{O1} O. Mustafa, Z. Algadhi, Eur. Phys. J. Plus \textbf{134}
(2019) 228.

\bibitem{O2} O. Mustafa, S. H. Mazharimousavi, Int. J. Theor. Phys.
\textbf{46} (2007) 1786.

\bibitem{O3} O. Mustafa, Phys. Lett. \textbf{A 384} (2020) 126265.

\bibitem{O4} O. von Roos, Phys. Rev. \textbf{B 27 }(1983) 7547.

\bibitem{O5} O. Mustafa, Ann. Phys. \textbf{440} (2022) 168857.

\bibitem{O6} O. Mustafa, Eur. Phys. J. C \textbf{82} (2022) 82.

\bibitem{O7} O. Mustafa, Ann. Phys. \textbf{446} (2022) 169124.

\bibitem{O8} O. Mustafa, Eur. Phys. J. Plus \textbf{138} (2023) 21.

\bibitem{O9} O. Mustafa, Phys. Lett. B \textbf{839} (2023) 137793.

\bibitem{O10} R. Shanker, Principles of Quantum Mechanics (Plenum
Press, NY and London, 7th Edition (1988)).

\bibitem{O11} A. Ronveaux, \textit{Heun's Differential Equations}
(Oxford University Press, New York, 1995).

\bibitem{key-012} M. Abramowitz and I. A. Stegun, Handbook of mathematical
functions with formulas, graphs, mathematical tables, Dover Publications,
New York, (1970).

\bibitem{key-013} A. Bouzenada, A. Boumali, Ann. Physics ,\textbf{452},169302(2023).


\end{thebibliography}
\end{document}